\documentclass[12pt]{iopart}
\usepackage{verbatim}
\usepackage{graphicx}

\usepackage{xspace}

\renewcommand{\Re}{\mathop{\mathrm{Re}}}
\newcommand{\MHi}{Mi\-chel\-son-in\-ter\-fe\-ro\-me\-ter}
\newcommand{\Mi}{Michelson interferometer}
\newcommand{\Pod}{Power-Recycled}
\newcommand{\gHw}{gravi\-ta\-tion\-al-wave}
\newcommand{\gw}{gravitational wave}
\newcommand{\mEx}[1]{\ensuremath\exp{\left(#1\right)}}
\newcommand{\eom}{electro-optic modulator}
\newcommand{\mEq}[1]{Equation~\ref{#1}}
\newcommand{\zr}{\ensuremath{z_{\rm R}}}
\newcommand{\Rr}{Rayleigh range} 
\newcommand{\mExB}[1]{\ensuremath\exp{\Bigl(#1\Bigr)}}
\newcommand{\mybs}{beam splitter}

\newcommand{\mFig}[1]{Figure~\ref{#1}}

\newcommand{\FINESSE}{\textsc{Finesse}\xspace}
\newcommand{\Finesse}{\textsc{Finesse}\xspace}

\newcommand{\w}{\ensuremath{\omega}}
\newcommand{\I}{\ensuremath{{\rm i}\,}}
\newcommand{\T}{\ensuremath{\,t}}
\newcommand{\shotH}{shot-noise}

\newcommand{\Geo}{GEO\,600}
\newcommand{\GEO}{GEO\,600}
\newcommand{\FI}{Faraday isolator}
\newcommand{\FP}{Fabry-P´erot}
\newcommand{\HG}{Hermite-Gauss}

\newcommand{\pd}{photo diode}

\newcommand{\mTab}[1]{Table~\ref{#1}}

\let \IG \includegraphics
\begin{document}
\title[Frequency domain interferometer simulation with higher-order 
spatial modes]{Frequency domain interferometer simulation with higher-order spatial modes}
\author{A~Freise\footnote[1]{To whom correspondence should be addressed 
}\ddag, G~Heinzel\S, H~L\"uck\S, R~Schilling\S, B~Willke\S\ and K~Danzmann\S}

\address{\ddag\ European Gravitational Observatory, Via E. Amaldi, 56021 Cascina (PI), Italy}
\address{\S\ Max-Planck-Institute for Gravitational Physics
(Albert-Einstein-Institute)
and University of Hannover, Callinstr. 38, D-30167 Hannover, Germany}
\begin{abstract}
\Finesse is a software simulation allowing one to compute the optical properties
of laser interferometers used by interferometric \gHw\ detectors 
today. This fast and versatile tool has already proven to be 
useful in the design and commissioning of \gHw\ detectors. 
The basic algorithm of \Finesse numerically computes the light
amplitudes inside an interferometer using \HG\ modes in the frequency domain.
In addition, \Finesse provides a number of commands for easily generating and plotting
the most common signals including power enhancement, 
error and control signals, transfer functions and \shotH -limited sensitivities.

Among the various simulation tools
available to the \gw\ community today, \Finesse
provides an advanced and versatile optical simulation
based on a general analysis of user-defined 
optical setups and is quick to install and easy to use.
\end{abstract}
\pacs{04.80.N, 95.55.Y, 07.60.L, 42.25.H}

\submitto{\CQG}

\ead{andreas.freise@ego-gw.it}

\maketitle

\section{Introduction}
The recently built interferometric \gHw\ detectors use a new class of laser interferometers:
new topologies have been formed combining well-known interferometer types such as Michelson or
\FP\ interferometers. In addition, the search for \gw s requires optical systems
with a very long baseline, a large circulating power and an extreme
stability. 
The properties of this new class of laser interferometers have been subject to 
extensive research.
\Finesse is an interferometer simulation that has been 
developed as a tool for designing and validating interferometer
topologies and, especially, the necessary control configurations\footnote{
By \emph{control configurations} we refer to the various possible methods
for stabilising a chosen optical setup to the desired operating point,
both for longitudinal and alignment degrees of freedom.}.
It is currently available in the Internet \cite{homepage}; the
package contains the program itself, several examples and detailed documentation 
(a graphical user interface for
\Finesse is available also \cite{luxor}). \Finesse is part of a collection
of simulation tools for advanced interferometry available to 
the gravitational wave community \cite{staic}.

The main task of the program lies in the computation of 
the amplitudes of light fields in an user-defined 
optical setup. This is done by transforming the user's description of the 
interferometer into a set of linear equations, which is then solved
numerically using a well-known algorithm for sparse matrices \cite{sparse}.
With the light amplitudes known at each location inside the interferometer,
the `optical problem' is completely solved. For fast
simulations, \Finesse provides many features for automatically generating
the most common interferometer output signals. 
A variety of output signals
can be computed including field amplitudes, light intensity and light power, 
optionally demodulated by up to 5 Fourier frequencies. 
In fact, almost every parameter of the interferometer description can be tuned during the
simulation. \Finesse automatically calls Gnuplot (a free graphics software \cite{gnuplot})
to create 2D or 3D plots of the output data.
An optional text output provides information about the
optical setup such as mode mismatch coefficients, eigen-modes of cavities and beam sizes.

\section{Mathematical description of interferometers}
Various techniques for computing the properties of optical systems exist.
One principal aim in developing \Finesse was to create a fast and accurate tool that
is easy to use, and techniques applied here were chosen accordingly.
At the
same time special approximations were to be excluded to provide an easy 
derivation of quantitative
error estimates and a tool that is as flexible and versatile as
possible. All interferometric detectors today employ optical systems with similar 
features: a) In normal operating condition, all parameters such as mirror positions, angles and
light powers should be as constant as possible (by virtue of sophisticated stabilisation systems).
b) The light fields inside the interferometer can be understood as a sum of the
laser input light at a fixed wavelength and different field components
with small frequency offsets (typically $1$\,Hz to $100$\,MHz).
c) The optical systems are designed such that the light
beams have a small divergence angle and are always small compared to the (main) optical
elements. 
Such systems can be well described using the frequency domain
and the paraxial approximation. Both methods are well-known, provide 
relatively intuitive results and were thus chosen as a baseline for \Finesse.
Special analysis tasks such as the lock acquisition 
process or non-linear couplings require a time-domain analysis. When
the paraxial approximation cannot be used, more general models for 
beam propagation, such as FFT propagation, have to be employed. Unfortunately, these
methods require more computing power than a paraxial analysis in the frequency domain.

The paraxial approach
is a straightforward extension to a plane-wave analysis;
for a better understanding the basic principles of \Finesse are explained 
in the following 
using plane waves only.
Further on, the necessary extensions for using higher-order 
transversal modes are explained.

\subsection{The plane-wave approximation}
In the plane-wave approximation the geometric properties transverse to 
the optical axis are ignored; thus the analysis becomes one-dimensional.
A light field at one point in space can be described as a sum
of frequency components:
\begin{equation}\label{eq:freqsum}
E(t)~=~\sum_{j}~a_{j}~\mEx{\I \w_j \T} 
\end{equation}
with $\w_j$ as the angular frequency of the light field 
and $a_j$ as complex coefficients, holding the amplitude and
phase information.
A common method for analysing optical systems is to define linear coupling 
coefficients for the (complex) amplitudes of 
the electric field component of the light field locally for
every optical component. For example, the coupling coefficients for a light
field (plane wave) at a plane, partly reflecting surface can be given as:

\vspace{.8cm}{
 \hspace{2cm}$
\left( 
\begin{array}{c}
\rm Out1 \\
\rm Out2 
\end{array}
\right)=\left(
\begin{array}{cc}
r  & \I t \\
\I t  & r \\
\end{array}\right)
\left(
\begin{array}{c}
\rm In1 \\
\rm In2 
\end{array}\right)
$}

\nopagebreak
\hspace{10cm}\IG [viewport=0 0 4 2, scale=0.4] {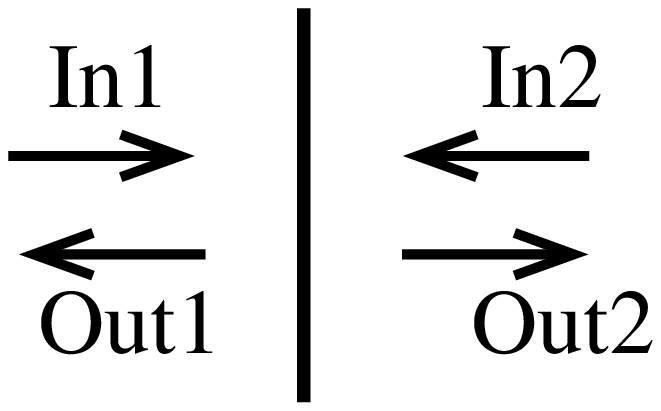} 
\vspace{.2cm}\\
with $r$ and $t$ as the amplitude coefficients for reflectance 
and transmittance
of the surface, `$\rm In$' as the incoming light fields and `$\rm Out$' the 
outgoing light fields\footnote{The phase change upon reflection can also be
implemented differently; this representation was chosen for its symmetry \cite{ghh}.}. 
In the plane-wave mode the interferometer can be built from
the following components: mirror, \mybs, free
propagation (`space') , input light (`laser'), \eom, \FI\
and \pd s.
All together, the linear couplings of the optical
components in a user-defined setup form a set of linear equations
for the light field amplitudes. It is solved numerically
by \Finesse to obtain the steady-state solution (light amplitude 
for every location in the optical system) from which the output
signals are generated with respect to the defined detectors.

\paragraph{Lengths and tunings:}
Interferometers often employ propagation of light fields over
large distances (e.g. meters to kilometers) while at the same time the
interference conditions depend on the microscopic differences
between two lengths (e.g. less than a micrometer). To 
be able to accurately describe the system on both scales, any
distance $D$ is given by two numbers \cite{ghh}: the length $L$, defined as 
the multiple of the laser wavelength $\lambda_0$ yielding the
smallest difference to $D$, and a microscopic
\emph{tuning} $\phi$ defined as the remaining difference between $L$ and $D$
($\phi$ is given in radians with $2\pi$ referring to one wavelength).
In \Finesse lengths are used as a parameter of free propagation,
whereas tunings are introduced as local parameters of mirrors and \mybs s.

\paragraph{Static response, frequency response:}
Any interferometer output signal can
be computed as a function of a quasi-static parameter
change (henceforth called \emph{static signal}) or
as a function of the Fourier frequency of a test
signal (henceforth referred to as \emph{transfer function}).
Whereas the static response can be used to search for
operating points with suitable control signals, the transfer functions
are commonly used to compute the sensitivity of the system and the 
coupling of various noise sources into the \gw\ signal.
In addition, transfer functions provide the necessary data for 
the design of the frequency response of the control system filters.

\paragraph{Modulation, demodulation:}
Modulation of a light field (in phase, amplitude or frequency) can be 
implemented by adding new field components, so-called sidebands, with a
frequency offset to the carrier equal to integer multiples of the modulation frequency.

The interferometer matrix has then to be solved sequentially for all existing frequency
components. In general, a large number of fields with 
different Fourier frequencies is present on each detector. 
Often the detected photo current is to be demodulated by one or more
frequencies and subsequently low pass filtered to generate the desired output signal.
In the frequency domain the demodulation (which, in the time domain, represents a 
multiplication with a periodic function, here a sine wave) plus the additional
low pass can be realized by a selection of the respective beat signals between the
field amplitudes. For example, a signal proportional to the photo current 
of a detector illuminated by
a field as given in \mEq{eq:freqsum} is:
\begin{equation}
S_0~=~|E|^2~=~\sum\limits_{i=0}^N\sum\limits_{j=0}^N a_ia_j^*~e^{\I(\w_i-\w_j)\T}
=:\sum\limits_{i=0}^N\sum\limits_{j=0}^N A_{ij}~e^{\I\w_{ij}\T}
\end{equation}
After a single demodulation at frequency $\w_x$ the low frequency component of the
signal can be written as:
\begin{equation}
S=\sum\limits_i\sum\limits_j \Re\left\{A_{ij}~e^{-\I\varphi_x}\right\}\quad\mbox{for}
\quad \{i,j~|~i,j\in\{0,\dots,N\}~\wedge~\w_{ij}=\w_x\}\\
\label{eq:single_demod}
\end{equation}
with $\varphi_x$ as the demodulation phase.

\subsection{The paraxial approximation: \HG\ modes}
The plane-wave approximation provides information
about longitudinal degrees of freedom of an interferometric
detector only. The alignment of the optical components, however,
plays an important role for the stability and noise performance
of an interferometer. Furthermore, the transverse properties of the
optical components often affect the longitudinal
signals also. 
These effects can be computed using \HG\ modes that are exact solutions
of the paraxial wave equation. Depending on the amount of deviation
(misalignment or mismatch), a number of higher-order modes describe
a paraxial system in a good approximation.
\Finesse offers two additional components in the Hermite-Gauss mode: the thin lens and
the beam analyser used to plot the shape of the beam on the
detector. In addition, attributes for mirrors and \mybs s can now be set, 
namely radius of curvature and angle of misalignment. 

\paragraph{Tracing the beam:}
In the paraxial approximation the electrical field can be 
described as:
\begin{equation}\label{eq:HG_intro1}
E(t,x,y,z)~=~\sum_{j}~\sum_{n,m}~a_{jnm}~u_{nm}(x,y,z)~\mEx{\I(\w_j \T -k_j z)} 
\end{equation}
with $u_{nm}(x,y,z)$ describing the spatial properties of the \HG\ modes and
$a_{jnm}$ as complex amplitude factors 
($k_j=\w_j/c$). Each set of \HG\ modes $u_{nm}$ represents a complete
set uniquely described by the so-called Gaussian beam parameter
$q_0=\I\zr-z_0$, with $\zr$ being the \Rr\ and $z_0$ the position of the beam waist.

In a general optical setup the shape of the beams is not known a priori but
depends on the resonance condition of cavities.
On the other hand, a set of base functions 
has to be defined before starting the calculation.
The choice of possible $q_0$ values is critical; even for only small
deviations of $q_0$ from the optimum beam parameter the number of higher-order
transverse
modes that have to be included in the simulation increases significantly.
To avoid numeric errors and to optimise the simulation speed the Gaussian
beam parameters should be set with great care.

A possible method for finding reasonable beam parameters for every
location in the interferometer (every \textit{node} in \FINESSE)
is to first set the beam parameters where they are known intuitively 
and then derive the remaining beam parameters from these initial ones:
The user can define the beam parameters directly for a given location
in the optical setup, especially for the laser outputs. 
In addition, \Finesse can compute the eigen-modes
of (geometrically stable) cavities and sets the respective beam parameters on every node
that is part of the respective cavity.
A beam-tracing algorithm is used to derive and set beam parameters for the 
remaining nodes.
\textit{Trace} in this context means that a beam starting at a node
with an already known beam parameter is propagated through the
optical system and the beam parameter is transformed according
to the optical elements passed using the ABCD matrix-formalism
\cite{siegman}.

If the interferometer is confined to a plane as in \FINESSE, it is convenient to use
two beam parameters: $q_{\rm s}$ for the sagittal plane and $q_{\rm t}$ for the tangential
plane so that the \HG\ modes can be written as:
\begin{equation}
u_{nm}(x,y,z,q_0)=u_n(x,z,q_{\rm t})~u_m(y,z,q_{\rm s})
\end{equation}
Also geometrical attributes such as radius of curvature or misalignment angles can
be given for the sagittal and tangential planes separately. This allows one to include
optical elements with astigmatism in the analysis.

\paragraph{Coupling coefficients:}
During free propagation the \HG\ modes are not coupled. Coupling is introduced in general whenever
the interference between two beams is to be computed. In that case one beam has to be
described in the base set of \HG\ modes of the other beam. This change of base system
introduces couplings between the TEM modes except for
the special case of perfect \emph{alignment} and \emph{matching}.
`Matching' refers to the difference in the Gaussian beam parameter
$q_0$ of the two beams. So-called coupling coefficients have to be computed for every 
mirror and \mybs. A set of \HG\ modes $u_{nm}$ characterised by the beam
parameter $q_0$ can be expanded using a different set of \HG\ modes
\cite{bayer}:
\begin{equation}
u_{n m}(q)\mExB{\I(\w t -k z)}=\sum_{n',m'}k_{n m n' m'}u_{n' m'}(q')\mExB{\I(\w t -k z')}
\end{equation}
where $u_{n'm'}$ represents \HG\ modes with a different beam
parameter $q_0'$ in a coordinate system $x',y',z'$ that is rotated around the $y$ axis
by the \emph{misalignment angle} $\gamma$.
The coupling coefficients are thus given by the following projection:
\begin{equation}\label{eq:tem_conv}
k_{n m n' m'}=\mEx{\I 2 k z' \sin^2\left(\frac{\gamma}{2}\right)}\int\!\!\!\int\!dx'dy'~
u_{n' m'}\mEx{\I k x' \sin{\gamma}}~u^*_{n m}
\end{equation}
For small misalignments ($\gamma \ll 1$) the 
coupling coefficients can be split into $k_{n m n' m'}\approx k_{n n'}k_{m m'}$.
In \cite{bayer} the above projection integral is partly solved
and the coupling coefficients are given by simple sums
as functions of $\gamma$ and the mode mismatch parameter $K$, which are 
defined as:
\begin{equation}
{\renewcommand{\arraystretch}{1.5}
\begin{array}{l}
K=\frac12(K_0+\I K_2)\ \ \mbox{with:}\\
K_0=(z_R-z_R')/z_R'\,,\qquad K_2=((z-z_0)-(z'-z_0'))/z_R'
\end{array}}
\end{equation}
\mTab{tab:coeff} shows the coupling coefficients for two special cases as an example.

\begin{table}[h]
\begin{minipage}{\textwidth}
{\small\hfill
\begin{tabular}{c|ccc}\small
 $k_{nn'}$ & 0& 1 & 2\\
\hline
0 & $ab$ & 0 & $\frac{1}{\sqrt{2}}ab^3K^*$ \\
1 & 0 & $a^3b^3$ & 0 \\
2 & $\frac{-1}{\sqrt{2}}ab^3K$  & 0 & $a^5b^5-\frac12 ab^5|K|^2$ \\
\end{tabular}\hfill
\begin{tabular}{c|ccc}
$k_{nn'}/d$  & 0 & 1 & 2\\
\hline
0 & $1$ & $-X^*$ & $0$\\
1 & $X$ & $1$ & $-\sqrt{2} X^*$\\
2 & $0$ & $\sqrt{2}X$ & $1$\\
\end{tabular}\hfill}

\caption[Coupling coefficients for misalignment]
{\label{tab:coeff}The above tables give the coupling coefficients for two
special cases and mode numbers $n,n'\leq2$: the left table refers to perfect alignment
but mismatch using
$a=\left(1+K_0\right)^{1/4}$ and
$b=\left(1+K^*\right)^{-1/2}$,
whereas the right table gives the coefficients for small misalignments
and no mismatch with
$X=-q\sin{(\gamma)}/w_0$ and
$d=\exp{(-|X|^2/2)}$ (terms of order $X^2$ or higher have been omitted 
in this example).
}
\end{minipage}
\end{table}

\paragraph{Detectors:}
The \HG\ modes (as they are used here) are ortho-normal. Therefore
the photo current upon detection on a single-element \pd\ (for simplicity
shown here for one frequency component only) is proportional to:
\begin{equation}\label{eq:hg_dc_det}
S=\sum_{n,m}a_{nm}a_{nm}^*
\end{equation}

More interesting for the \HG\ modes are different detector types that are
sensitive to the shape of the beam, such as split 
photo detectors. \Finesse can be used with such detectors
of arbitrary design by defining the \emph{beat coefficients}.
For an arbitrary split detector the respective signal is computed as:
\begin{equation}\label{eq:hg_dc_det}
S=\sum_{n,m}\sum_{n',m'} c_{nmn'm'} a_{nm}a^*_{n'm'}
\end{equation}
with $c$ as the beam coefficient matrix. For example, a detector split
in the $x$ direction is usually set up to detect the difference between the
two halves thus being sensitive only to the beat signals of symmetric with
asymmetric modes (in the $x$-direction). The coefficients thus are:
\begin{equation}
c_{nmn'm'}=\left\{
\begin{array}{l}
1\,\,  \mbox{when}\,\, (n+1) + (n'+1)\,\, \mbox{is odd and}\,\,m=m'\\
0\,\,  \mbox{otherwise}
\end{array}
\right.
\end{equation}
The subsequent demodulation of the signal is performed exactly as
in the plane-wave mode. Using split detectors the control
signals for automatic alignment systems or other 
similar geometrical control systems can be computed.

\section{Results and validation}
Being a versatile simulation tool, \Finesse is based on a complex
code both with respect to the programming itself and to the mathematical 
description of the physics. Careful testing is necessary 
to gain a high level of confidence with respect to the
correctness of the results.

\Finesse has been subject to many implicit and explicit tests during its development
and use: The plane-wave analysis with \Finesse was compared against a number
of algebraic calculations of simple interferometers and cavities.
Such tests should provide a good validation since 
the program is not tailored to any special topology or
configuration. In order to test the \HG\ extensions, the simulation
results had to be compared to other optical simulations: The 
propagation of Gaussian beam and the functioning of the beam trace algorithm
could be verified using OptoCad \cite{optocad}. Coupling coefficients
generated by direct numerical integration were used to verify 
and validate the actual implementation derived from \cite{bayer}.
Finally, complex examples of analyses with higher-order modes 
were compared to the result of an FFT propagation code.

In addition, the simulated data obtained with \Finesse have been
compared successfully with experimental data during the commissioning
of \Geo\ \cite{bfw}. On its own, such a comparison does not represent 
a sufficient test but confirms the dedicated tests mentioned above.
The following example shows one of the first employments of 
\Finesse to understand data from the \Pod\ \Mi\ of \Geo.
\begin{figure}[p]
\begin{minipage}{\textwidth}
\begin{center}
\IG [bb=50 50 320 300,clip,scale=.61] {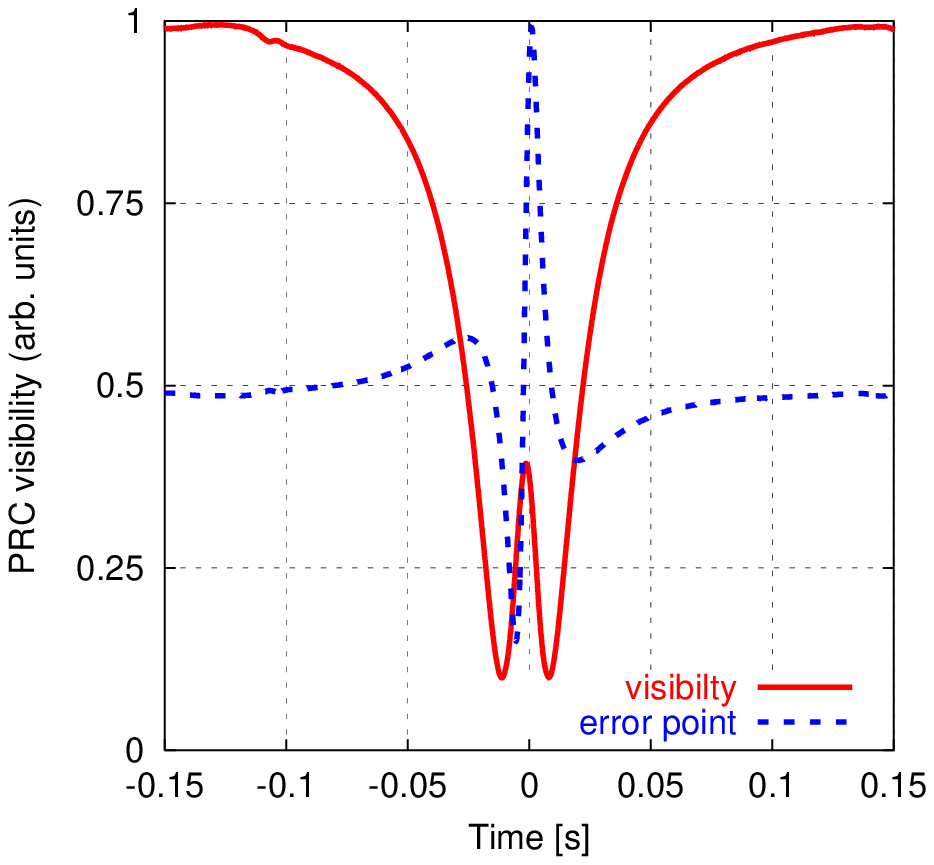} 
\IG [bb=165 10 295 245,scale=.5] {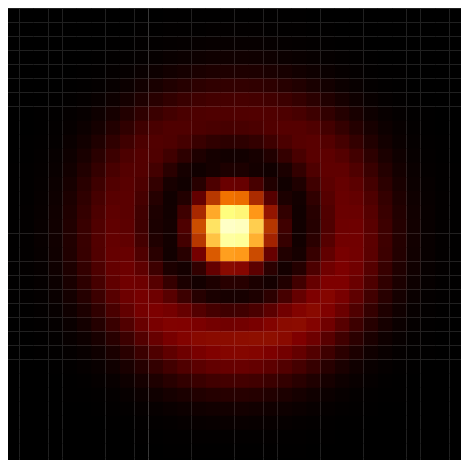} 
\IG [bb=50 50 350 300, clip, scale=.61] {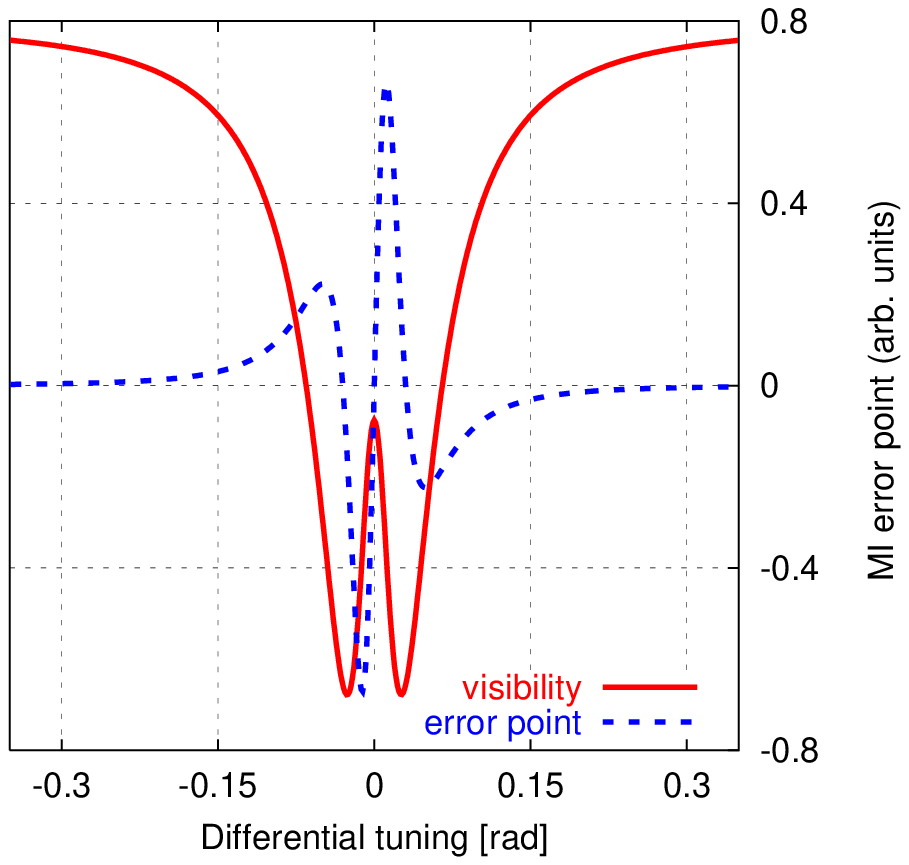} \\
\IG [bb=50 50 320 300,clip,scale=.61] {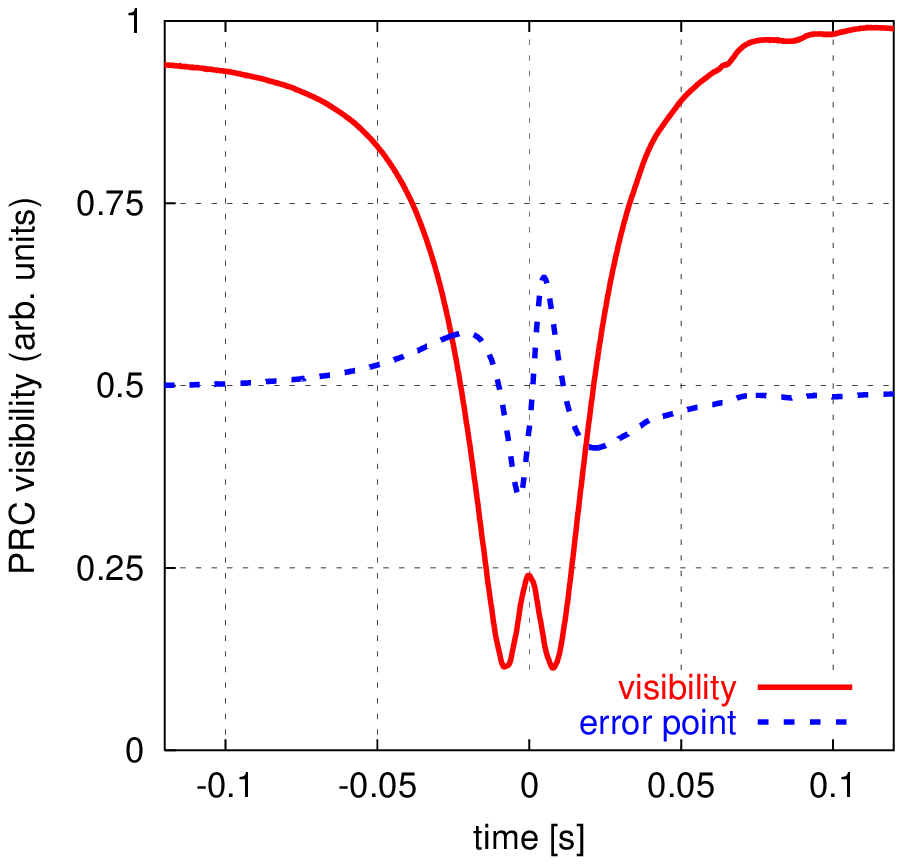} 
\IG [bb=165 10 295 245,scale=.5] {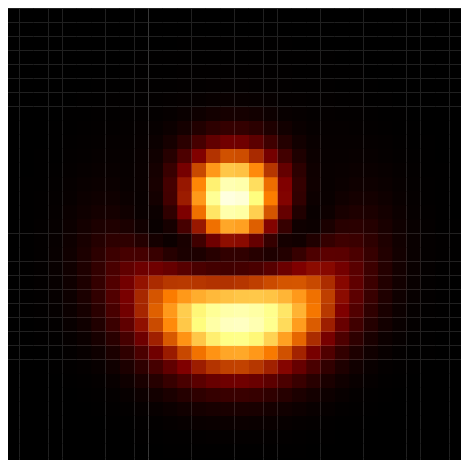} 
\IG [bb=50 50 350 300, clip, scale=.61] {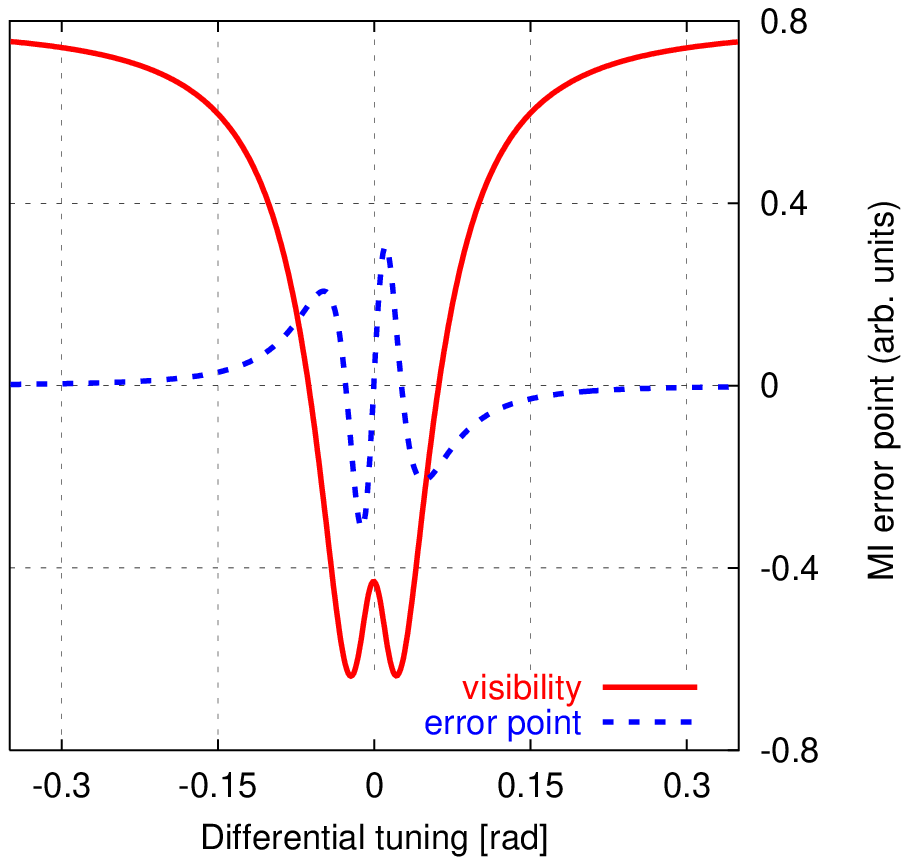} \\
\IG [bb=50 50 320 300,clip,scale=.61] {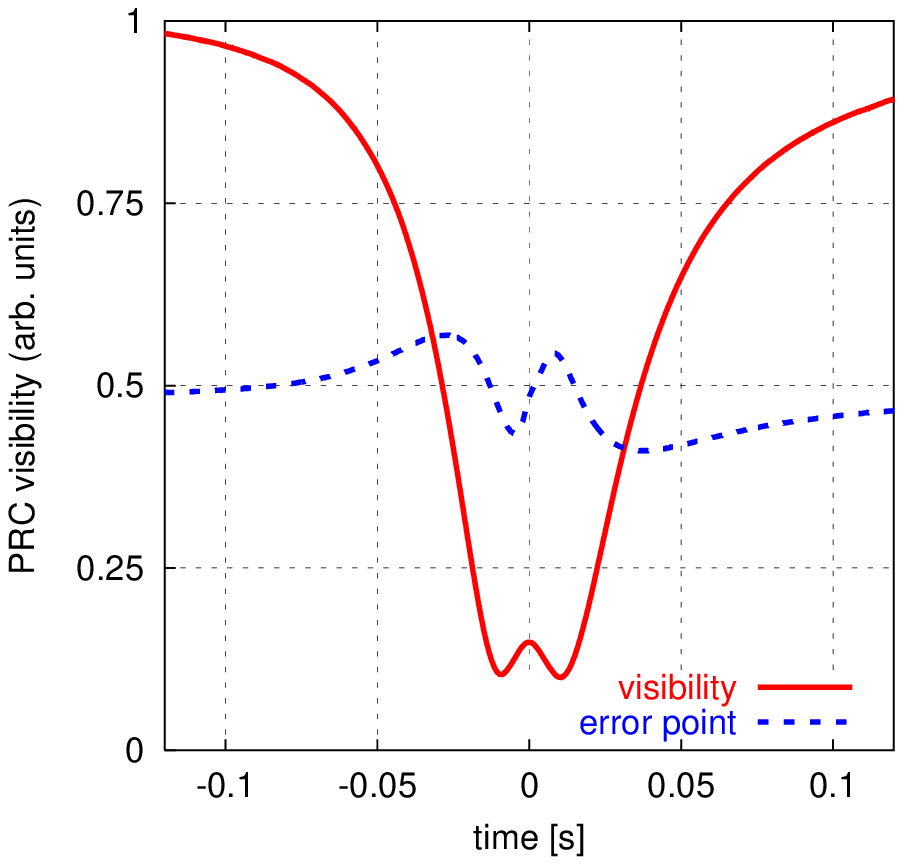} 
\IG [bb=165 10 295 245,scale=.5] {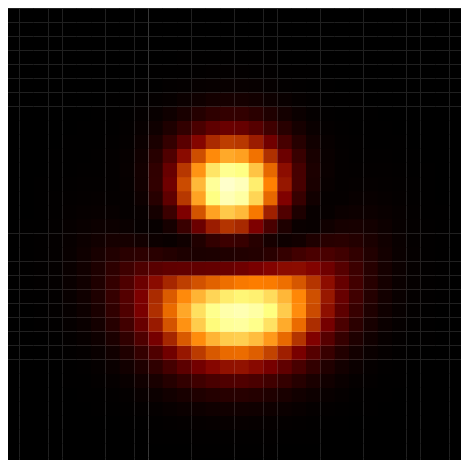} 
\IG [bb=50 50 350 300, clip, scale=.61] {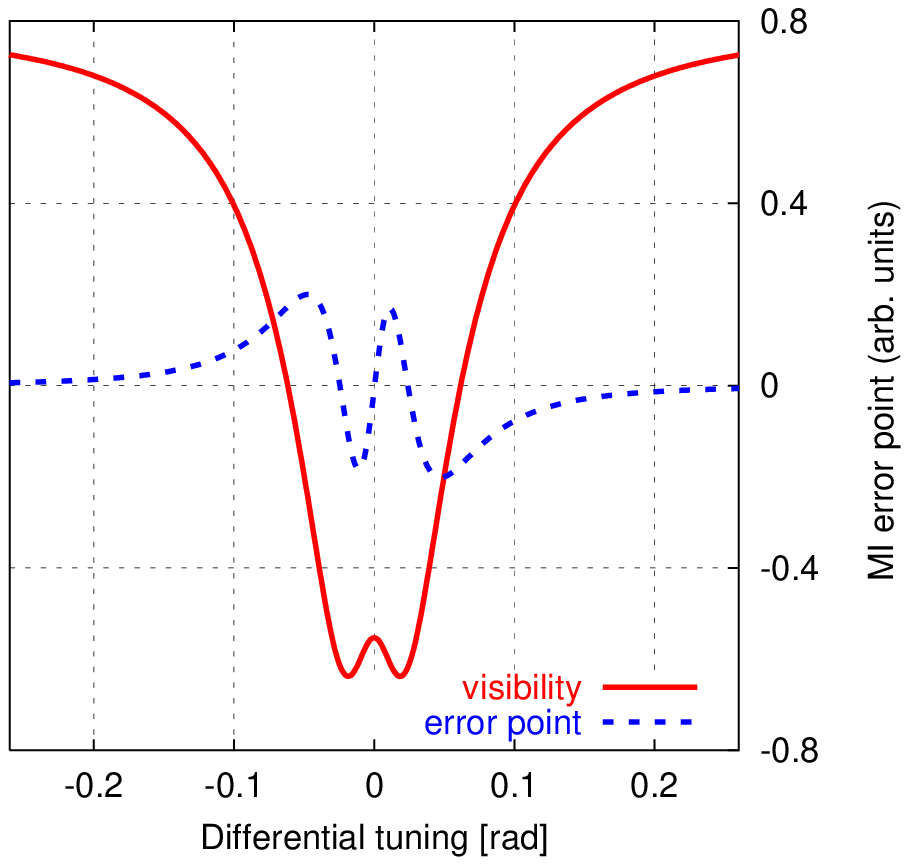} 
\end{center}
\end{minipage}
\caption[Measurement and simulation of \MHi\ error signals]{\label{fig:mi_fringe}
Measurement and simulation of \MHi\ error signals: The left column shows measured signals for three typical 
fringes of the \Mi ; each graph
shows the MI error point and the PRC visibility (the fraction of the light power 
reflected by the cavity) while the \Mi\ passes through a dark fringe. The
right column shows respective signals obtained by a \FINESSE simulation.
From top to bottom, the
simulations were done for increasing misalignments of the end mirrors 
(0.2\,$\mu$rad, 3\,$\mu$rad and 4\,$\mu$rad). The center column shows the 
corresponding dark fringe pattern computed by the simulation.}
\end{figure}
\mFig{fig:mi_fringe} shows measured and computed signals for the \Pod\ \Mi :
The \Pod\ cavity (PRC) was locked and the
\Mi\ (MI) was freely passing through dark fringes. Zero time
indicates the center of the respective dark fringe. 
The left column shows three typical measured fringes; each 
graph shows the \MHi\ error signal and the PRC visibility, respectively. It can be
seen that the size of the center part of both signals strongly
varies, whereas the signal shape far away from the center remains similar. The `double dip' in
the PRC visibility is due to the changing reflectance of the \Mi : It 
is really highly reflective only at the
dark fringe so that the PRC becomes over-coupled
as designed. If the \Mi\ is moving away from the dark fringe, its reflectance is
decreased. At some point the PRC becomes impedance-matched where the visibility
is minimised. By further decreasing the MI reflectance, the visibility increases;
the cavity becomes under-coupled. Thus, 
a full fringe of the \Mi\ results in a double structure, as seen in \mFig{fig:mi_fringe},
with the minima indicating the impedance-matched states.

Simulations with \FINESSE have been used to understand the
signal shapes. The right column in \mFig{fig:mi_fringe} shows the computed signals.
The parameters of the simulations were adjusted as follows:
The general input file describing the \GEO\ optical setup includes measured parameters
if available and nominal parameters otherwise. 
To obtain a computed dark-fringe pattern similar to the observed ones,
the radius of curvatures of the end mirrors (which were not known accurately)
were adjusted accordingly in the simulation. 
The computed MI error point and the PRC visibility signals
matched the measured signals with the strong center very well 
(top graph in \mFig{fig:mi_fringe}).
It turned out that the form of the computed signals could be matched to the 
measured signals by introducing a misalignment of the end mirrors
into the simulation (\mFig{fig:mi_fringe} middle and bottom).
Both deviations were later determined by further measurements:
The radii of curvature of the end mirrors deviated from the
specification as predicted by the simulation, and the varying central shape of the 
signal at the dark fringe was caused by 
an unexpected small suppression of the tilt movement 
of the \mybs, which has a similar effect on the output signals as a differential tilt movement  
of the end mirrors.

The simulations helped to identify and characterise these defects and
showed that the \Mi\ otherwise worked as expected. In addition,
it became clear that a good and stable alignment of the
interferometer mirrors is necessary for obtaining a proper error signal. 

\Finesse has been widely used in several projects; however, it has been most
frequently utilised during the commissioning of \Geo.
Some of these simulation results can be found in this issue \cite{hal,icm,hrg,malik}
and in \cite{adf}. 

\ack
The program has been and still is constantly extended and improved. Many people
helped improving the code by testing it and sending comments.
Especially, the authors would like to thank Kenneth Strain, 
Guido M\"uller, Keita Kawabe, Jan Harms and the \Geo\ team
for their support and encouragement. 
We would like to acknowledge support also from the DFG (Deutsche Forschungsgemeinschaft).


\section*{References}
\begin{thebibliography}{99}
\bibitem{homepage}
 Freise A 1999 {\it Finesse} http://www.rzg.mpg.de/\char126 adf
\bibitem{luxor}
 Harms~J 2002 {\it Luxor} http://www.aei.mpg.de/\char126 jah
\bibitem{staic} http://www.phys.ufl.edu/LIGO/LIGO/STAIC.html
\bibitem{sparse}
 Kundert~K~S and Sangiovanni-Vincentelli~A 1988 University of California, Berkeley
\bibitem{gnuplot} Williams~T and Kelley~C {\it Gnuplot} http://www.gnuplot.info
\bibitem{ghh} Heinzel~G 1999 Ph.D.~Thesis, University of Hannover
\bibitem{siegman}
 Siegman~A~E 1986 {\it Lasers} University Science Books, Mill Valley
\bibitem{bayer} Bayer-Helms~F 1984 {\it Appl. Opt.} {\bf 23} 1369--1380
\bibitem{optocad} Schilling~R 2002 {\it OptoCad} internal note
\bibitem{bfw} Willke~B \etal \CQG this issue
\bibitem{hal} L\"uck~H \etal \CQG this issue
\bibitem{icm} Malec~M \etal \CQG this issue
\bibitem{hrg} Grote~G \etal \CQG this issue
\bibitem{malik} Rakhmanov~M \etal \CQG this issue
\bibitem{adf} Freise~A 2003 Ph.D.~Thesis, University of Hannover\hfill\newline 
http://www.amps.uni-hannover.de/dissertationen/freise\_diss.pdf
\end{thebibliography}
\end{document}